\documentclass[iop]{emulateapj}
\shorttitle{Resolved stars in IZw18} 
\shortauthors{Contreras et al.}

\begin{document}

\title{Spatial distribution and evolution of the stellar populations and
candidate star clusters in the Blue Compact Dwarf I Zwicky 18\footnote{Based 
on observations with the NASA/ESA Hubble Space Telescope, obtained at the 
Space Telescope Science Institute, which is operated by AURA Inc., under 
NASA contract NAS 5-26555.}}

\author{R. Contreras Ramos, F. Annibali, G. Fiorentino, M. Tosi}
\affil{INAF-Osservatorio Astronomico di  Bologna, via Ranzani 1, 40127, 
Bologna, Italy}
\email{rodrigo.contreras@oabo.inaf.it}
\author{A. Aloisi}
\affil{Space Telescope Science Institute, 3700 San Martin Drive, Baltimore,
MD21218}
\author{G. Clementini}
\affil{INAF-Osservatorio Astronomico di  Bologna, via Ranzani 1, 40127, 
Bologna, Italy}
\author{M. Marconi, I. Musella}
\affil{INAF-Osservatorio Astronomico di Capodimonte, via Moiariello 16, 
80131, Napoli, Italy}
\author{A. Saha}
\affil{National Optical Astronomy Observatory, P.O. Box 26732, Tucson, AZ 
85726}
\and
\author{R. P. van der Marel}
\affil{Space Telescope Science Institute, 3700 San Martin Drive, Baltimore,
MD21218}

\begin{abstract}
The evolutionary properties and spatial distribution of I~Zwicky~18 stellar 
populations are analyzed by means of HST/ACS deep and accurate photometry.  
The comparison of the resulting Colour-Magnitude diagrams with stellar 
evolution models indicates that stars of all ages are present in all the 
system components, including objects possibly up to 13 Gyr old,
intermediate age stars and very young ones.  
The Colour-Magnitude diagrams show evidence of thermally pulsing asymptotic 
giant branch and carbon stars. Classical and ultra-long period Cepheids, as 
well as long period variables have been measured.
About 20 objects could be unresolved star clusters, and are mostly concentrated
in the North-West (NW) portion of the Main Body (MB). 
If interpreted with simple stellar population models, these objects indicate a particularly 
active star formation over the past hundred Myr in IZw~18.

The stellar spatial distribution shows that  the younger ones are more centrally 
concentrated, while old and intermediate age stars are distributed 
homogeneously over the two bodies, although more easily detectable at the 
system periphery. 
The oldest stars are best visible in the Secondary Body (SB) and in the South 
East (SE) portion of the MB, where crowding is less severe, but 
are present also in the rest of the MB, although measured with larger 
uncertainties.
The youngest stars are a few Myr old, are located predominantly in the MB and 
mostly concentrated in its NW portion. The SE portion of the MB 
appears to be in a similar, but not as young evolutionary stage as the NW, 
while the SB stars are older than at least 10 Myr. There is then a sequence 
of decreasing age of the younger stars from the Secondary Body  to the SE 
portion of the MB  to the NW portion.

All our results suggest that IZw18 is not atypical compared to other BCDs.
\end{abstract}


\section{Introduction}
I Zwicky 18 (also Mrk 116 or UGCA 166, hereafter IZw18) is the most 
intriguing Blue Compact Dwarf (BCD) galaxy in the local Universe. 
It became famous right after its discovery,
when \cite{searle72} measured from its emission-line spectrum an 
oxygen abundance [O/H]= --1.14 dex, only 7$\%$ of the solar value, 
indicating an almost unprocessed gas content. The first studies 
on its color and composition \citep{sargent70,searle72,searle73}
emphasized  its almost primordial evolutionary status. 
IZw18 shows very blue colors, U$-$B$=-0.88$ and 
B$-$V$=-0.03$ \citep{vanzee98}, suggesting the presence of a very young 
stellar population, with a current star formation rate (SFR) much higher 
than the past mean value. The total mass of IZw18 from the rotation curve at a radius of 
10$^{\prime\prime}\div$12$^{\prime\prime}$ is estimated  to be
$\sim 10^8$ M$_{\odot}$ 
\citep[e.g.,][]{davidson85,petrosian97,vanzee98}, where the neutral gas
corresponds to $\sim$70$\%$ of the total mass, but only $ 10^7$
M$_{\odot}$ of HI is associated with the optical part of the galaxy 
\citep[e.g.,][]{lequeux80,vanzee98}.

Subsequent spectroscopic studies in IZw18 \citep{lequeux79,davidson85,
dufour88,pagel92,skillman93,kunth94,stasinska96,garnett97,izotov98} 
have confirmed its extreme metal deficiency around 
1/30$\div$1/50 of Z$_{\odot}$. Nowadays IZw18 remains the star forming 
galaxy with the second lowest metallicity and the lowest helium 
content known \citep{izotov09}. This makes the system a fundamental point 
in the derivation of the primordial helium abundance 
\citep{izotov94,olive97,izotov98} and 
in the study of the properties of chemically unevolved galaxies. 

When discovered by Zwicky (1966), IZw18 was described as ``two galaxies 
separated by 5$^{\prime\prime}$.6 and interconnected by a narrow luminous 
bridge'', surrounded by two ``very faint flares'' at 24$^{\prime\prime}$ 
northwest. Subsequent CCD ground--based images \citep{davidson89,dufour90} 
revealed a more complex structure: the {\it two galaxies} turned out to be 
two star--forming regions of the same galaxy (usually indicated as NW and SE 
components), while the {\it two flares} are just the most prominent of a few 
nebulosities surrounding IZw18. These minor systems are roughly aligned 
toward the northwest and were initially considered at the same distance,
but subsequent spectroscopic studies showed that only one component
\citep[referred to as component C in][]{davidson89} is at the same distance  
as IZw18 and is physically associated with the main body 
\citep{dufour96a,petrosian97,vanzee98}. 
The other diffuse objects have been recognized as background galaxies 
\citep[see, e.g.,][]{dufour96b}. 
In the following we will refer to IZw18 and to component C respectively 
as main body (MB) and secondary body (SB) of IZw18.
Both bodies have been resolved into single stars for the first time 
by HST with the Wide Field Planetary Camera 2 (WFPC2) by \cite{hunter95} 
and \cite{dufour96b}.  

The observational features described above triggered the key question over
the nature of IZw18: is it 
a young galaxy which is presently experiencing its first burst of star 
formation, or is it an old system which has already formed stars in the past?
HST studies by different authors have tried to characterize
the evolutionary status of IZw18, with rather discordant
results, but with a clear trend intriguingly reminiscent of Oscar Wilde's 
Dorian Gray's
picture: the more one looks at it, the older it appears. 
\cite{aloisi99}, 
using HST/WFPC2 data, were the first to detect
Asymptotic Giant Branch (AGB) stars in the galaxy, and thus to
demonstrate that IZw18 is as old as at least several hundred Myr, 
and possibly much older.
\cite{izotov04} failed to detect red giant branch (RGB) stars 
with HST observations made by the Advanced Camera for Surveys (ACS),
and thus concluded that the galaxy is at most 500 Myr old. However, 
both \cite{momany05} and \cite{tosi07}, from independent re-analyses of 
the same ACS data set, suggested that IZw18 should be older than at 
least 1-2 Gyr, since it did appear to contain also Red Giant 
Branch (RGB) stars.
Our own time-series HST/ACS photometry 
\citep[GO 10586,][hereafter Al07]{aloisi07},
has allowed us to shed light on the situation, thanks to 
the combined information from
the color-magnitude diagram (CMD) and the lightcurves of the 
Cepheids we were able to measure. 

In fact, Al07 not only characterized the RGB and identified 
its Tip (TRGB) at $I_0 = 27.27 \pm 0.14$ mag, but, more importantly, 
detected for the first time a few Cepheids whose lightcurves allowed to
independently and more reliably pin down the galaxy's distance to the couple 
Mpc accuracy (Fiorentino et al. 2010, hereafter F10; Marconi et al. 2010).
The distance to IZw18 was estimated from the Cepheids using
theoretical $V,I$ Wesenheit relations specifically computed for the
very low metal abundance of the galaxy, and by theoretically fitting
the observed lightcurve of the most reliable Classical Cepheid with period 
8.71 days.  The Wesenheit relations provide an intrinsic distance 
modulus for IZw18 of $(m-M)_0=31.4 \pm 0.2$  (D=19.0$^{+1.8}_{-1.7}$ Mpc) 
adopting a canonical mass - luminosity relation, and of 
31.2 $\pm$0.2  (D=$17.4^{+1.6}_{-1.6}$ Mpc) when an
overluminosity of 0.25 dex is assumed for each
given mass (non canonical scenario).  The theoretical 
modeling of the 8.71 days Cepheid lightcurve 
provides 31.4 $\pm$0.1  (D=19.0$^{+0.9}_{-0.9}$ Mpc).
The TRGB identified by Al07 implies an instrinsic distance
modulus of 31.30 $\pm$ 0.17, in excellent agreement with the
Cepheid-based estimates. This agreement, in turn, provides independent support
to our interpretation of the CMD faint red sequence as an RGB.

Hence, IZw18 is actually  farther away than originally thought 
\citep[for instance][adopted from the literature a 
mean value of D=10 Mpc]{aloisi99} and is
in practice at slightly larger distance than the Virgo cluster of galaxies. 
This made it impossible in the past to resolve its older, hence 
fainter, stars, and quite challenging
even with the outstanding  performances of HST's current instrumentation.
Today IZw18 is one of the most distant system ever resolved into stars.

In this work, we have used the photometric catalogues of our previous works, which have been focused on determining the age (Al07) and the distance through the variable stars (F10).
In this paper we describe the evolutionary properties and the spatial
distribution of the resolved stellar populations and candidate star clusters 
of the main and secondary bodies of IZw18.
In Section 2, we briefly summarize the main differences in the photometric procedures adopted by Al07 and F10. The CMDs and the spatial distribution of stars in different evolutionary phases 
are presented  in Section 3 and 4, respectively. The results are analysed in 
Section 5 and summarized  in Section 6.

\section{The photometric data--set}

The data acquisition and reduction have been extensively described by Al07 
and F10. Here we only  summarize their main aspects. 
Time series photometry of IZw18 in the F606W (broad $V$) 
and F814W ($I$) filters was obtained with the Wide Field Channel (WFC) of the ACS  
in 13 different epochs properly spread over a time interval of 96 days (program Id. 10586, PI Aloisi). 
These proprietary data were complemented by 5 F555W ($V$) and 3 F814W archival ACS/WFC images
of IZw18 from HST (program Id. 9400, PI Thuan). 

IZw18 is quite distant, plenty of bright stars hiding and/or contaminating the fainter 
ones, and plenty of gas, which complicates the photometry of 
individual stars. In addition, the
throughput of the F606W filter is significantly affected by the $H_{\alpha}$
contribution of ionized gas (quite abundant in the MB of IZw18) and makes the
star detection and measurement more complicated. 
Thus, depending on the goal, one can find more appropriate to restrict the analysis of the galaxy populations to 
safer but fewer objects (Al07), or to more uncertain but numerous ones (F10).

In Al07, to produce the galaxy CMD, we built four
coadded images by stacking all single epoch images per filter for both
archival and proprietary data, separately. Then we performed PSF
photometry of each of these four coadded images using the
DAOPHOT/ALLSTAR package \citep{stetson87}.  The calibration to the 
Johnson-Cousins photometric system was
obtained using the transformations given by \cite{sirianni05}. 
Since the goal of that paper was to demonstrate that faint red stars 
actually existed where none had previously been
detected, the published CMD was produced following a strategy aimed at
minimizing the number of false detections. As a result, it contained only the
2099 sources which were independently detected
in all the four deep images (the $V$s and $I$s of both datasets), with good
match in the position (within one pixel) and magnitude 
(within $3.5\sigma$) of each source in the different images, and with 
sharpness $|sharp|\leq$0.5.  These are 
the most reliable objects measured in each of IZw18's bodies.  

In the case of F10, to produce the lightcurves of the candidate variables, we made use of
DAOPHOTII/ALLSTAR/ALLFRAME \citep[][1994]{stetson87} to get
PSF photometry on each single epoch (15 $I$ and 18 $V$ images, in
total between proprietary and archival data). To study the variability we relaxed the Al07 selection criteria, since variable stars are more easily distinguishable from spurious objects thanks to their brightness variation in the images at different epochs. Moreover, to collect a good sample of candidate variables at all magnitudes, we examined objects close to the detection threshold. 
This photometry resulted in a $I, V-I$ CMD of IZw18 
which contains 6859 sources with photometric errors
$\sigma_{V,I} < 0.5$ mag and $sharp$ parameter in the range from $-1$ to $+1$, 
and reaches $V\sim$ 30.0, i.e. more than half a magnitude deeper than  
Al07's photometry. The calibration to the Johnson-Cousins photometric system was
obtained using the transformations given by 
\cite{sirianni05}, as for the other data set. 

As shown above, there is a remarkable difference in the number of objects selected with the two approaches. 
To infer the properties of the stellar populations inhabiting IZw18 
minimizing the effects of uncertain photometry, we will base our results 
on the hyper-selected Al07 catalogue. Nonetheless, we consider it useful 
to show some features of the larger F10 catalogue both for completeness 
and for comparison.

\begin{figure}
\includegraphics[width=8.5cm]{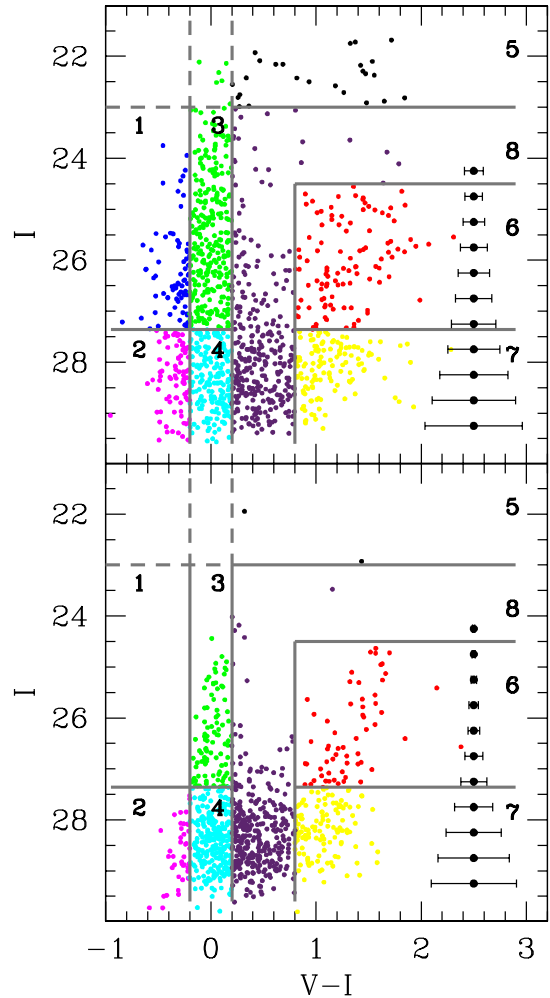}
\caption[]{CMD of the Al07 hyper-selected stars of the main (upper panel) and 
secondary (lower panel) body showing the 8 evolutionary 
zones described in the text. Notice that zones 3 and 5 partially overlap: this
is intentional to let the brightest stars of the blue plume be considered 
either as upper MS objects or as supergiants or as star clusters.} 
\label{cmdmap07}
\end{figure}

\begin{figure*}
\centering
\includegraphics[width=17cm]{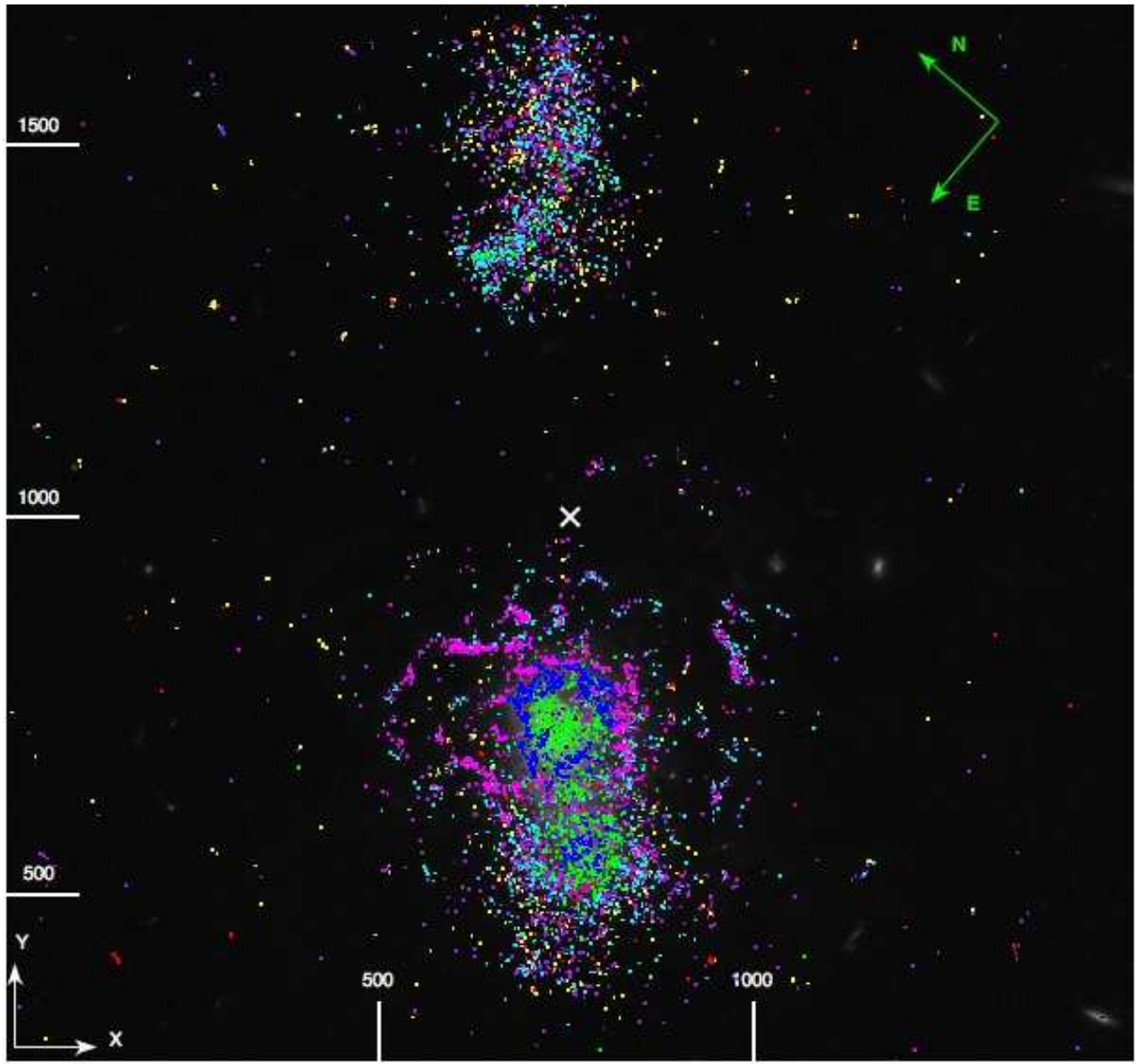}
\caption[]{Portion of one F814W image of IZw18,  with overimposed the 6859 stars measured 
by F10. The X-Y coordinates are in pixels (pixel scale $\sim 0.035^{\prime \prime}/pix$). 
The white cross indicates the center
of this portion ACS/WFC image, which has RA$=$09$^h$ 34$^m$ 01$^s$, 
DEC$=$55$^{\circ}$ 14$^{\prime}$ 33.3$^{\prime \prime}$.
The stars are divided and color-coded according the 8 CMD evolutionary zones
in Fig.~\ref{cmdmap07}.}
\label{image814}
\end{figure*}

Since our aim is to analyse the evolutionary properties of IZw18's stellar
populations, we have put special care in verifying whether the selection 
differences affect some CMD regions more than others. We have hence divided 
the CMDs of the two sets in 8 zones roughly representative of different 
evolutionary phases. 
Since the TRGB was detected by Al07 at $I=27.3$, we have adopted this 
magnitude as the border between bright and faint stars: 

\begin{itemize}
\item zone 1, stars in the extremely hot portion of the blue 
plume with $V-I\leq-0.2$ and brighter than $I=27.3$; 
\item zone 2: stars in the extremely hot portion of the blue 
plume with $V-I\leq-0.2$ and fainter than $I=27.3$; 
\item zone 3: stars in the ``standard" blue plume, 
with $-0.2<V-I\leq+0.2$, and brighter than $I=27.3$; 
\item zone 4: stars in the ``standard" blue plume, 
with $-0.2<V-I\leq+0.2$, and fainter than $I=27.3$; 
\item zone 5: objects with $I\leq23$  and any colour; 
\item zone 6: stars with $V-I\geq0.8$ and $24.5\leq I \leq 27.3$;
\item zone 7: stars with $V-I\geq 0.8$ and $I\geq 27.3$;
\item zone 8: all the remaining ones;. 
\end{itemize}
The number of objects measured in 
each zone is reported in Table~1 for both bodies and both data sets.

\begin{deluxetable}{lrrrrrrrr}
\tablecaption{Number of MB and SB objects in the Al07 and F10 data sets, and 
corresponding percentages, in the 8 CMD zones defined in the text and shown 
in Figure~\ref{cmdmap07}.}
\footnote{Notice that the total number of stars in the MB is
smaller than the sum of the values in the corresponding column, because the objects falling in
the overlapping zones 3 and 5 are counted twice:19 in F10 and 6 in Al07.} 
\tablewidth{0pt}
\tablehead{CMD  &N$_{MB} $&\% &N$_{MB}$ &\% &N$_{SB}$ &\%&N$_{SB}$ &\%  \\zone&(Al07)&&(F10)&&(Al07)&&(F10)\\
}
\startdata
zone 1 & 76  &6   &649  &13  &0   &0    &0    &0\\
zone 2 & 84  &7   &917  &20  &43  &5    &108  &5\\
zone 3 & 277 &23  &820  &17  &93  &9    &98   &4\\
zone 4 & 186 &16  &707  &15  &262 &29   &584  &28\\
zone 5 & 34  &3   &51   &1   &2   &$<$1 &2    &$<$1\\
zone 6 & 105 &8   &210  &4   &60  &6    &110  &5\\
zone 7 & 126 &11  &349  &8   &120 &13   &375  &18\\
zone 8 & 305 &26  &1060 &22  &332 &36   &838  &40\\
total  &1187 &100 &4744 &100 &912 &100  &2115 &100\\
\enddata
\end{deluxetable}

The zones are displayed on the Al07 CMDs  in Figure~\ref{cmdmap07}. 
Figure~\ref{image814}  shows an F814W image of IZw18 
with the F10 measured objects colour-coded according to the 8 CMD zones 
of Figure~\ref{cmdmap07}. We arbitrarily define as MB objects those with Y 
coordinate in the image smaller than 1200 px and as SB objects those 
with Y$\geq$1200 px. In Figure~\ref{cmdmap07}, and all the following ones, 
the CMDs of the two bodies are splitted following this definition.

We have found that the two photometries agree 
well with each other, once we take the different selection 
criteria and photometric depth into account.
The objects measured by both Al07 and F10 fall consistently 
in the same CMD zone (i.e. have the same colours and magnitudes).  

The F10 objects not present in the Al07 hyper-selected catalogue fall mostly 
in the fainter portions of the CMDs, as expected since they can have larger 
photometric error. An intriguing difference is that the MB zones 1 and 2 are 
much more 
populated in F10 than in Al07 (see Table~1), with a ratio N(F10)/N(Al07)=8.5 
in zone 1 and 10.9 in zone 2, significantly higher than the total MB ratio 
4744/1187=4.0. Figure~\ref{image814}  reveals that most of the 
F10 objects of zones 1, 2 and 4 fall on the gas-rich regions in and around
the MB. A possible cause for the discrepancy is therefore gas 
contamination in the F606W flux, even if in principle all photometric 
packages should overcome this kind of
effects by subtracting the surrounding background flux to each source. 

As a matter of fact, many of the objects in zone 1 and 2 show a discrepancy in $V$-magnitude depending on the used filter, e.g. F606W or F555W. If, we restrict the selection criteria on the F10 catalogue and keep only the objects with $\sigma_{V,I}<$0.1, most of those in the CMD zone 2 and several of those 
in  zone 1 are rejected.
We thus consider as real stars on the gas filaments only the 
brighter ones with smaller photometric error.

Finally, we have individually and visually inspected the brighter objects of  
F10 zone 1 using appropriate routines, and found that some of them do appear 
as real stars.
We consider the vast majority of the blue objects measured by F10 on the gas 
filaments and not present in the Al07 catalogue as likely spurious detection 
due to gas peaks, but  we do believe that some of the brighter 
zone 1 detections are real stellar objects.

To be conservative, we will base all our results on the restricted Al07
catalogue.

\section{Colour-magnitude diagrams}

\vspace{0cm}
\begin{figure*}
\centering
\includegraphics[width=15cm]{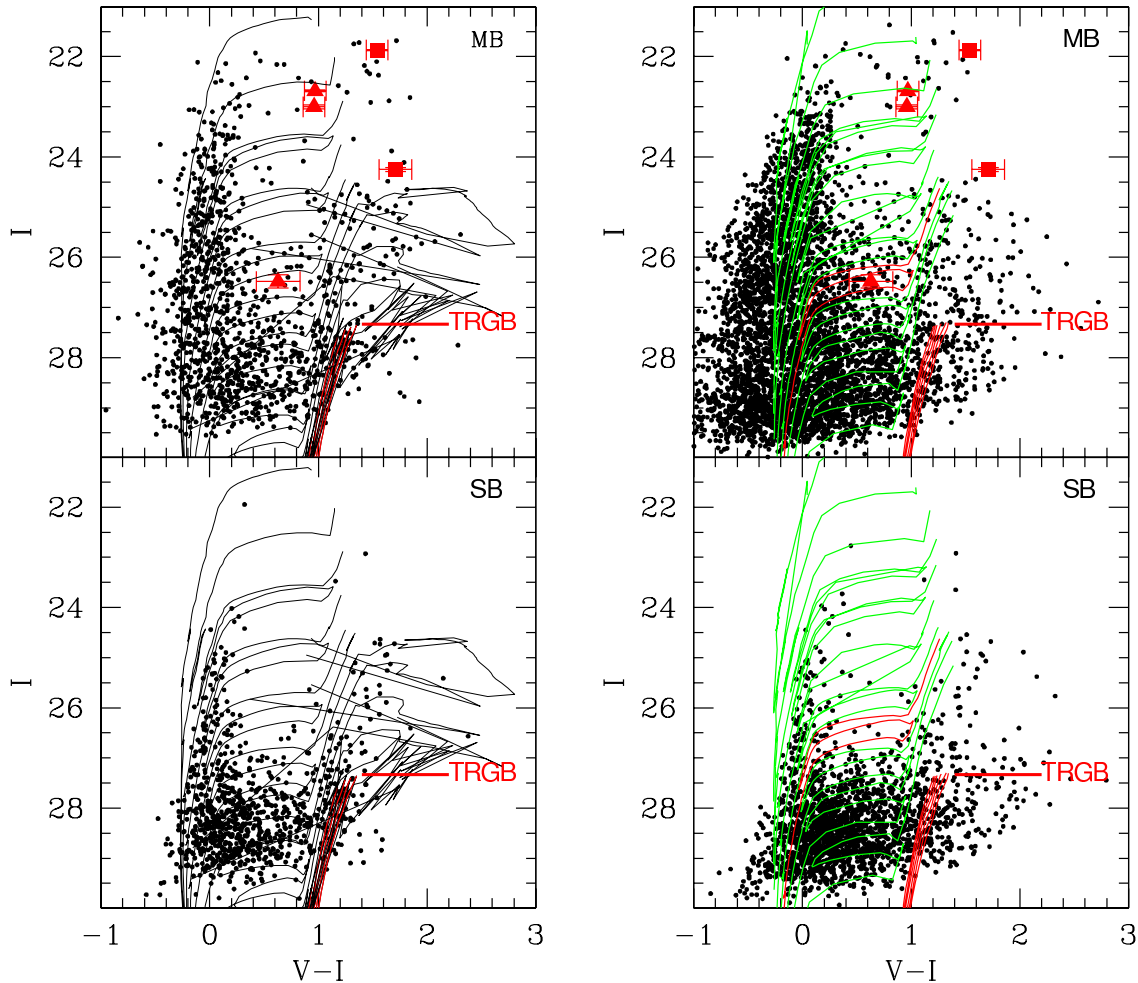}
\caption[] {CMD of IZw18 MB (top panels) and SB (bottom panels) obtained by 
Al07 (left 
panels) with restrictive selection criteria and by F10 (right panels) with 
less restrictive ones (see text).  The CMDs are calibrated in the 
Johnson-Cousins photometric system. 
The three confirmed Cepheids are highlighted by filled 
(red) triangles, and the two Long Period Variables by filled (red) squares. 
All confirmed variables are plotted according to their intensity-averaged mean
magnitudes and colors from F10. In all panels, stellar evolutionary models with
metallicity Z=0.0004 are overimposed to the data adopting $(m-M)_0$=31.4 
and E(B-V)=0.032. 
In the left panels we plot isochrones with detailed TP-AGB modelling 
\citep{girardi10}, and log(age) 6.75, 7.00, 7.25, 7.50 ... 
9.75, 10.00 (from top to bottom). 
The tangled, messy portions of the isochrones above the RGB correspond
to the TP-AGB unstable phases. In the right panels we plot the classical 
evolutionary tracks by \cite{fagotto94}. In green (in the electronic 
version) are the tracks for 3, 4, 5, 7, 9, 12, 15, 20, 30 and 60 M$_{\odot}$, 
and in red the 6 M$_{\odot}$ one, which best fits the 8.71d Cepheid. 
The TP-AGB phase is not covered by these tracks. The RGB portion of both the isochrones and evolutionary tracks are shown in red in the electronic version.}

\label{cmdisotracks}
\vspace{0cm}
\end{figure*}

To let the reader easily figure out the involved stellar ages, the left
panel of Figure~\ref{cmdisotracks} shows the CMDs of the hyper-selected 
Al07 objects with overimposed theoretical isochrones \citep{girardi10}. 
The isochrones have metallicity Z=0.0004 and are plotted 
adopting $(m-M)_0$=31.4 (Al07 and F10) and E(B-V)=0.032 
\citep{vanzee98}.
The  CMDs corresponding to the less restrictively selected F10
photometry are displayed in the right panel of Figure~\ref{cmdisotracks}, 
where stellar evolution tracks  are also plotted to let the reader 
visualize  what the involved star masses are. 
The plotted tracks, again with Z=0.0004, are the 
classical Padova models by \cite{fagotto94}.
We note that IZw18 is located at high galactic latitude ($b=+45^{\circ}$)
and therefore its CMDs do not suffer from significant contamination
from foreground stars belonging to our Galaxy, meaning that is not
necessary to correct for it our observed diagrams.

IZw18 is a starforming dwarf and its CMDs show all the features typical 
of these galaxies: a prominent blue plume with mean $V-I\simeq-0.1$ and a 
fainter and broad red plume with $V-I\geq0.8$. Once
compared with the stellar evolution tracks of Figure~\ref{cmdisotracks}, the 
blue plume turns out to correspond to high- and intermediate-mass 
stars in the main sequence (MS) phase or at the blue edge of the blue loops 
covered during central He burning. These two phases are resolved by HST 
photometry in nearby late-type galaxies, but at
the distance of IZw18 the photometric error inevitably is large enough to 
let them merge in just one broad sequence. 
Stars with colours intermediate between the blue and
the red plume are either supergiants (brighter than $I\simeq23$) or
intermediate-mass stars in the blue loop phase. The brighter ones might 
also be unresolved star clusters.

The red plume is populated by red supergiants ($I \leq 23$), 
AGB ($24.5 \leq I < 27.3$) and RGB stars ($I\geq27.3$). Our data allow 
to cover the brighter 1.5 magnitude portion of the RGB, where 
$\sigma_{Daophot}$ is $\pm$0.1 mag in $I$ and  $\pm$0.2 in $V-I$ (Al07) and 
the photometric error, estimated from extensive artificial star 
tests \citep{annibali11}, 
is smaller than $\pm$0.3 mag in $I$ and  $\pm$0.4 in $V-I$
(see Figure~\ref{cmdmap07}). The colour extension 
of the RGB is larger than this error and reflects, as usual, both the spreads 
in age and in metallicity of the resolved stars. However, given the extremely 
low metallicity of IZw18, in this case it is mostly the age that contributes 
to the RGB thickness. In other words, the reddest RGB stars are most likely 
low mass stars $10$--$13$ Gyr old. 

The redder red plume stars with $I \simeq 26.4$ are presumably carbon stars,
not too different from those measured in the Magellanic Clouds by 
\cite{vandermarel01}. The redder objects at $I \simeq 25.5$ are most 
probably stars of $4$--$5$ M$_{\odot}$ in their thermal pulsing (TP) phase 
at the end of the AGB, since their location 
in the CMD is perfectly consistent with the predictions of 
Marigo \& Girardi's (2007) TP-AGB models. 
These authors show in fact that at 
very low metallicities (Z $ \leq 0.001$) in (relatively) massive
intermediate-mass stars efficient nuclear burning at the base 
of the convective envelope may favour the formation of C-rich models and make 
their tracks describe pronounced excursions towards lower effective
temperatures. These excursions are recognizable in Figure~\ref{cmdisotracks} in
the tangled, messy portions of the isochrones above the TRGB.

\vspace{0cm}
\begin{figure}
\includegraphics[width=8.5cm]{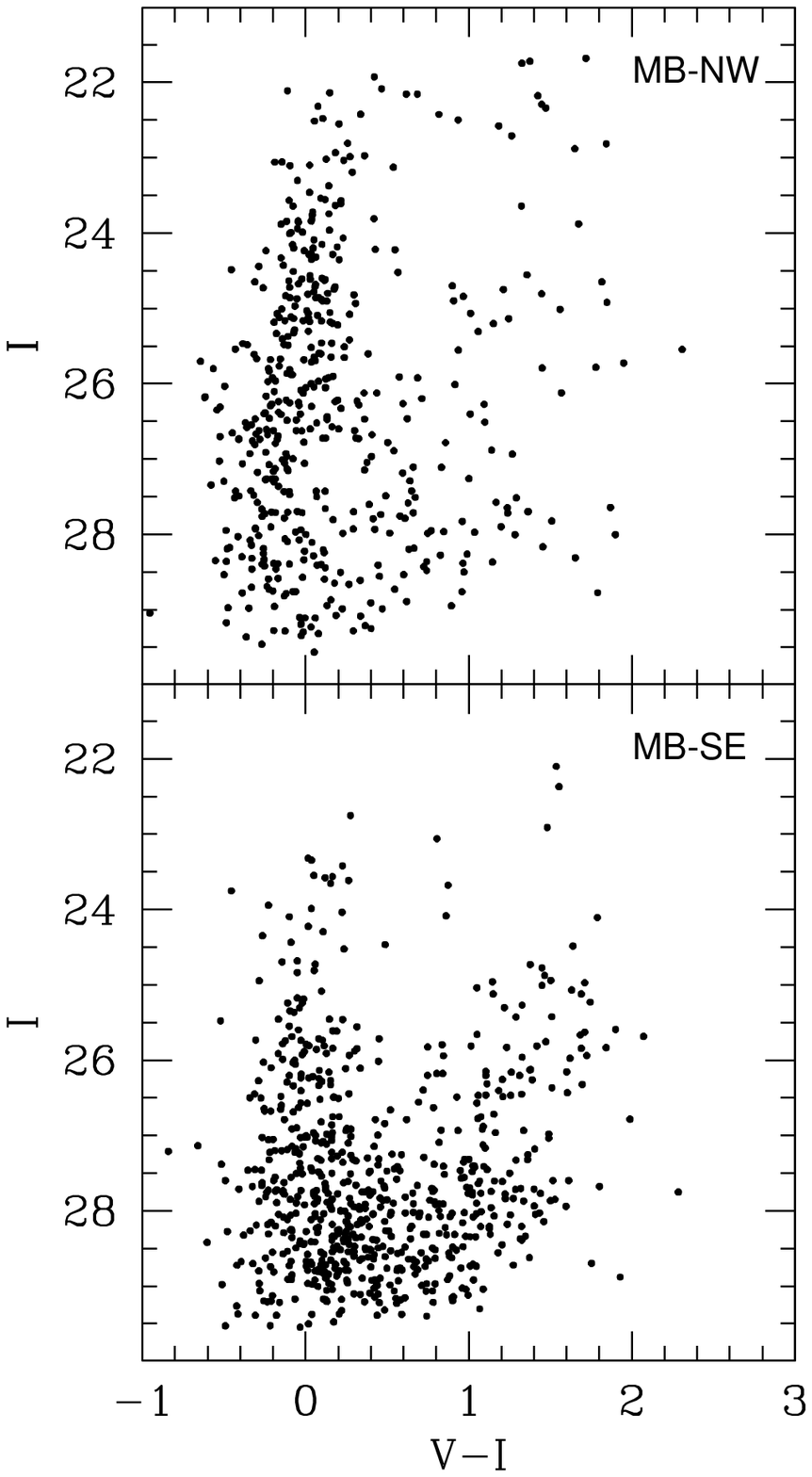}
\caption[]{CMD of the MB objects in the hyper-selected Al07 catalogue, 
separating those in the NW region (Y$>$650 px, top panel) from those in the 
SE region (Y$\leq$650 px, bottom panel).  There are 478 objects in the NW 
region and 704 in the SE one.}
\label{mb_separate}
\vspace{0cm}
\end{figure}

Not only does the MB, by definition, host a larger number of stars; it 
also contains a larger fraction of young, blue and bright stars than the 
SB. Both the luminosity and the number of the brighter stars in the
SB are significantly lower than in the MB. Almost no supergiant is
found in the SB and no Cepheid. Moreover, while the blue plume of the MB
covers the entire range of blue massive stellar evolution tracks and youngest
isochrones, that of the SB turns off the MS around $I\simeq28$, with apparently
no bright star younger than 10 Myr being detected there.
 This is a significant difference, which 
reflects, as discussed already by \cite{aloisi99}, both a different rate
and a different history of star formation,
with the MB being more active than the other at recent epochs, and 
the SB more active earlier than about 100 Myr ago, as suggested by the 
higher density of objects around the isochrones older than this age in 
the left-bottom panel of Figure~\ref{cmdisotracks}. 
At the faintest magnitudes, we 
should also consider that crowding affects the MB more severely, 
preventing the detection of its faintest 
stars. This may be particularly the case in the NW part of the MB.

The variable stars in IZw18 have been identified by F10
exploiting our ACS time-series photometry. The intensity-averaged mean
magnitudes and colors of the confirmed Cepheids are plotted as 
filled (red) triangles in Figure~\ref{cmdisotracks}, while those of the
confirmed Long Period Variables (LPVs) are plotted as filled (red) squares. 
Their position (see F10 for identifications and 
nomenclature)  in the CMD indicates that the two bright ultra-long period 
Cepheids (ULPs) V1 and V15  are supergiants, unless 
they are hosted by unresolved star clusters making them appear brighter. 
The Classical Cepheid V6 with period P=8.71d is on the blue loop and is 
best fitted by a $\sim$6~M$_{\odot}$ star 
\citep[see also F10 and][]{marconi10}. 
The  brighter LPV V4 is a supergiant, while the
other confirmed LPV V7 lies at the bright edge of the CMD region 
of the TP-AGB stars and is presumably a Mira. 

For a finer analysis of the evolutionary properties of IZw18 components, 
we have further divided the Al07 objects according to their position in 
the MB. Figure~\ref{mb_separate} shows the CMDs of the objects with 
Y coordinate in the image larger (top panel) or lower (bottom panel) than 
650 px (see Figure~\ref{image814}). This in practice separates objects in 
the NW (top) part of the MB from those in the SE (bottom). The two 
CMDs are fairly similar to each other, although the NW one contains more 
objects in zone 5 and is more dispersed than the other. In particular, 
the red plume of the SE region is much better defined than that of the 
NW and allows to clearly distinguish the RGB and its Tip from the AGB.
The larger spread of the NW CMD is likely due to the presence  of a 
larger number of bright objects and thick ionized gas filaments that  
make the measurements of fainter (and older) stars more uncertain. It is 
plausible that the gas ionization  is itself due to the massive 
luminous objects populating the NW region; so, they may represent the 
only real difference between the northern and southern parts of the MB.   
We will further discuss these objects in the next section.

In spite of the very different number of objects selected in the two data sets,
once the different depths of the resulting CMDs (Figure \ref{cmdisotracks}) 
are taken into account, their stellar evolutionary 
phases turn out to be populated in similar proportions, both in the MB and 
in the SB (see Table~1). As discussed in the previous section, the only 
significant difference is that the MB blue plume is much more populated, 
bluer and thicker in the F10 CMD. 

\bigskip

\section {Spatial distribution and evolutionary properties of IZw18 
resolved stars}

We have analysed the spatial location of the various types of stars
populating IZw18 and checked if stars of different ages are evenly or unevenly
distributed. To this purpose, we have used the CMD zones of 
Figure \ref{cmdmap07}. Thanks to the comparison with the
evolutionary tracks, we can assign to each zone the following evolutionary
properties: 
\begin{itemize}
\item zone 1 corresponds to massive and intermediate mass stars on the MS , younger than $\sim$ 10--15 Myr;
\item zone 2 to fainter intermediate mass stars on the MS, younger than$\sim$ 70 Myr;
\item zone 3 to massive and intermediate mass stars on the MS or at the blue edge of the blue loop phase, younger than $\sim$ 70 Myr;
\item zone 4 to fainter intermediate mass stars on the MS or at the blue edge of the blue loop phase, younger than $\sim$ 300 Myr;
\item zone 5 to supergiants, younger than $\sim$ 10 Myr and, possibly, to unresolved star clusters; 
\item zone 6 to AGB stars, ages from $\sim$ 35 to 300 Myr;
\item zone 7 to RGB stars, ages from $\sim$ 1 to 13 Gyr;
\item zone 8 in most cases corresponds to intermediate-mass stars in the blue 
loops, but also to some supergiants, ages from $\sim$ 15 to 300 Myr. 
\end{itemize}

\begin{figure*}
\includegraphics[width=18cm]{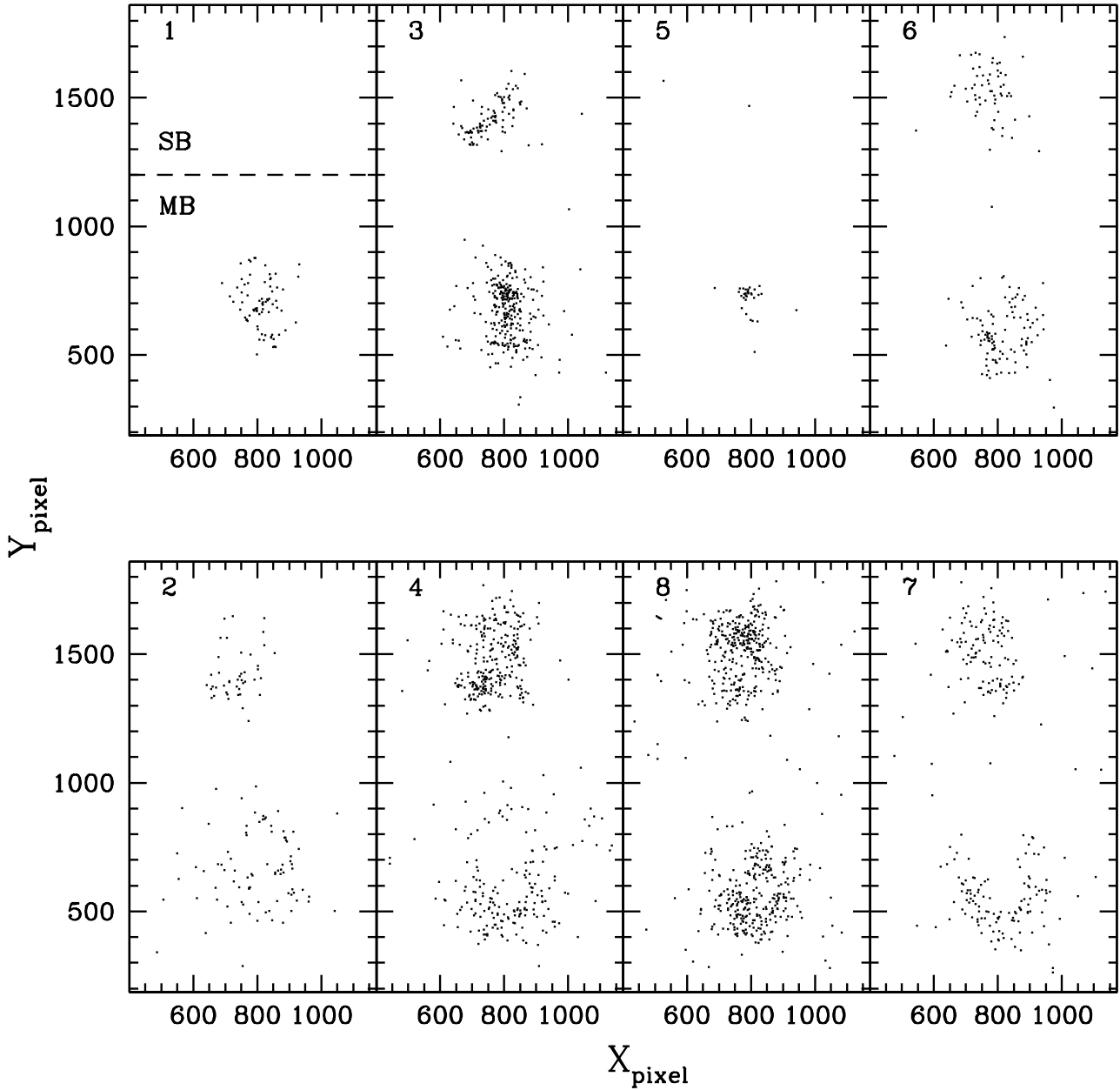}
\caption[]{X-Y map of the Al07 hyper-selected objects of Figure~\ref{cmdisotracks}, 
splitted according to the 8 CMD evolutionary zones (see text and 
Fig.~\ref{cmdmap07}). Objects with Y$\geq$1200 are associated to the SB, 
the others to the MB.  In the latter, the NW portion 
corresponds to Y$\geq$650 and the SE portion to Y$<$650.}
\label{map07}
\end{figure*}

The spatial distribution in the X-Y map of IZw18 of the Al07 objects
measured in the two bodies and falling in each of the 8 CMD zones is 
reported in Figure~\ref{map07}. 

This map shows that the SB does not contain stars in zone 1. In other 
words the SB has 
had very little, if any, star formation activity in the last few Myr. 
Zone 1 stars covering the whole range of magnitudes (hence with masses from 15--20
M$_{\odot}$ up) are numerous in the MB and show a sort of 8-shaped
spatial distribution. If we look at the fainter portion of the extremely blue
stars (zone 2), we find that both bodies contain several of these objects and
that in the MB they seem to delineate the same 8-shape as their brighter
counterpart.

The bright stars in zone 3 have masses above 
6 M$_{\odot}$  and are also present in both bodies. Their spatial 
distribution in Figure~\ref{map07} forms an 
irregular ellipsoidal clump of objects with significant central concentration 
(the MB) with a comma-shaped companion (the secondary: Zwicky's ``flare").
This is the overall familiar appearance of IZw18, since
these are the stars more easily visible, although not resolvable from
ground. Once again, the SB stars of zone 3
don't reach the luminosity of the brightest ones hosted in the
MB (see Figure~\ref{cmdmap07}) : while the whole mass and luminosity 
ranges are covered in the latter, only stars less luminous than $I\simeq24$ 
exist in the former. Taking into
account that any Initial Mass Function disfavours the most massive stars, we
consider it more likely that the brighter blue stars in the SB 
are 12--15 M$_{\odot}$ stars at the blue edge of the blue loop rather 
than more massive stars. In this case, the star formation activity in 
the secondary must have been quite low, or absent, in the last 10 Myr.

Fainter (lower mass) blue plume stars (zone 4) can be either MS stars with
masses between 6--20 M$_{\odot}$ or stars at the blue edge of the blue loop with mass
in the range 3--5 M$_{\odot}$ (see the tracks in right panel of 
Figure~\ref{cmdisotracks}).  
They are present in both bodies, and
are more abundant in the SB than in the MB, where they are not
found in the central regions and have a spatial distribution complementary to
that of their brighter counterpart of zone 3. This is clearly due to the severe
crowding affecting the MB and, particularly, its inner regions, where the
many bright stars completely hide fainter ones. In the SB the recent
star formation activity has been less intense, crowding is less severe, and
faint stars are much more easily detectable.

The lower activity of the SB at recent epochs is also confirmed by the
objects of zone 5. As apparent from Figure~\ref{map07}, we find only 2 such 
bright objects in the SB (one of which is quite far away and may
actually be a background system), and 31 in the MB.  The latter are fairly confined in the inner
regions, and most often in the NW portion of the MB. They significantly 
contribute to the severe crowding affecting the central area. If interpreted 
in terms of individual stars, these objects are
supergiants and confirm that the central region of the MB has experienced 
a strong star forming activity in the last 10 Myr or so. 

Admittedly, some of these very bright objects might be not individual 
stars but unresolved star clusters. We discuss this possibility in 
Section~\ref{cluster_candidates}.

The objects falling in the CMD zone 6 are 
(see the tracks in Figure~\ref{cmdisotracks}) AGB stars with mass lower 
than 9 M$_{\odot}$, and age (see the isochrones in the same Figure) in
principle between 35 Myr and 13 Gyr, but in practice peaked at, at least, 
several hundreds Myr, once the stellar lifetimes and phase durations are 
taken into account. They include the TP-AGB and carbon stars with 
4-5 M$_{\odot}$ (horizontal finger at $I\simeq$25.5 in the CMDs) predicted 
by \cite{marigo07} for these low metallicities 
and the carbon stars at $I\simeq$26.4 discussed by Al07.
Zone 6 objects are equally present and evenly distributed in both bodies,
although missed in the most crowded portions of the MB.

The RGB stars are in zone 7: objects with mass below 1.8-2.0 M$_{\odot}$ 
and age between 1 and $\sim13$ Gyr. The RGB is notoriously affected by the 
age-metallicity degeneracy, i.e. the ambiguity that stars with the same 
brightness and colour can be older and metal poorer or younger and metal 
richer. Hence, without independent spectroscopic metallicity measurements 
one cannot assign the age to RGB stars simply from their position in 
the CMD. For this reason the
photometric identification of the RGB normally puts only a lower limit of
1-2 Gyr (i.e. the minimum time required by low-mass stars to reach that
phase) to the age of the examined stellar population. However, in the
case of IZw18, its extremely low metallicity somewhat relaxes the effect of the
age-metallicity degeneracy, since the redder RGBs cannot be metal rich
objects and therefore must be old. In these conditions, the
colour extension of IZw18 RGB, which is  broader than the
photometric error and covers the whole range of isochrone ages,
can be taken as a signature of the presence of stars possibly as old as 13 Gyr.
RGB stars are faint and therefore their detection is strongly
affected by crowding and photometric performances. This is why only with 
the ACS has it been finally possible to resolve and measure them: not even HST 
was able to reach the RGB, with the less sensitive WFPC2. The red giants are 
evenly distributed in the SB and visible only in the
outer regions of the MB, due to the excessive crowding of its inner parts. It
is interesting to notice that the MB RGBs are located along a semi-arc
in the system periphery. We ascribe this asymmetry to the presence of thick gas
filaments in the regions where RGBs are not detected: overlapping filaments can
clearly hide such faint objects. We thus believe that RGB stars can actually
be evenly distributed also in the MB.

Finally, zone 8 collects all the remaining objects, with a large range of 
brightness and colour, and different ages and evolutionary phases. Those 
with $I\geq$24.5 are stars in the 3--10 M$_{\odot}$ mass range
in the central helium burning phase and with ages between 30 and 300 Myr. The
brighter ones are 10--15 M$_{\odot}$ mass stars in any post-MS evolutionary 
phase, with age of the order of 15--30 Myr. These mixed objects are evenly 
distributed in both bodies, with the only exception of the two most 
crowded NW and SE spots of the MB, where they are missing. Interestingly, 
like the objects of zones 2, 4 and 7,
they are found in similar numbers in the two bodies. Even considering 
the effect of the different crowding, this suggests that as far as low 
and intermediate mass stars are concerned, the SB is not significantly less 
populated than the MB.

As for the variable stars, the two ULPs V1 and V15 are located in the NW 
portion of the MB, while the classical Cepheid V6 and the two LPVs V4 and 
V7 are in the SE portion (see F10 figure 5). 
Interestingly, all the confirmed variables belong to
the MB: of the 34 good candidate variables examined by F10, almost
half (15) were actually detected in, or around, the SB, but for none of them 
satisfactory lightcurves and periods were derived. Ground-based
observations of the brighter variables are being performed to better
characterize their pulsation and evolutionary properties. In particular, 
further data on the ULPs will allow us to verify if they are the extension 
to higher masses of Classical Cepheids and, in case, to calibrate them as 
extremely useful primary indicators of cosmological interest.

Finally, we emphasize that no star is found in the region between the 
MB and the SB, either in the Al07 or the F10 catalogues. That region looks 
completely transparent, with only background galaxies being detected there. 
This shows that the two bodies are completely separated. They are members 
of a two-body system, but have no star connecting each other. If tidal 
interactions are at work, they don't appreciably affect the stellar 
populations of the two components.

\subsection{Cluster candidates \label{cluster_candidates}}

At 19 Mpc, ACS/WFC photometry cannot 
recognize round regularly shaped clusters whose sizes fall within the Point 
Spread Function. In this section we consider the possibility that zone 5 objects are actually 
not individual stars but unresolved star clusters, and discuss the implications on IZw18's evolutionary status.

\begin{figure}
\includegraphics[width=8cm]{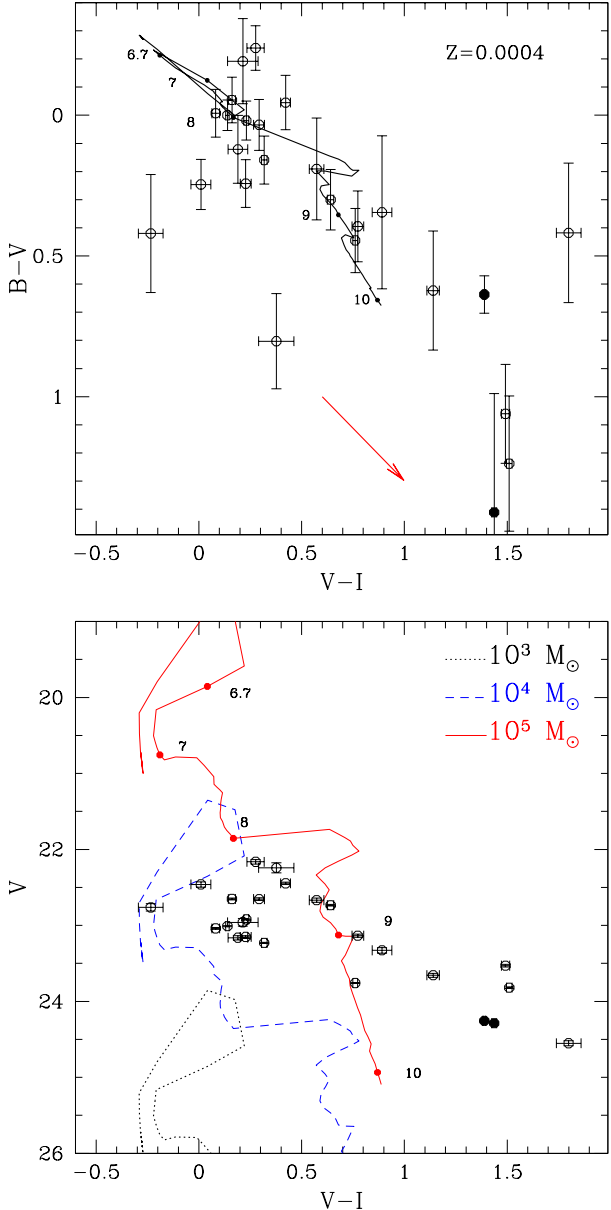}
\caption[]{Johnson-Cousins $B-V$ versus $V-I$ color-color diagram (top) 
and $V$ versus $V-I$ color-magnitude diagram (bottom)
for the objects in zone 5 (cluster candidates) that have been cross-correlated with the WFPC2 $(B,V)$ 
photometry presented by \cite{aloisi99}. Full symbols denote objects 
located far away from the MB and SB which could be
background galaxies. Overplotted are the Padova simple stellar populations
\citep{girardi10} for a Z$=$0.0004 metallicity, ages from 1 Myr 
to $\sim$13 Gyr, and masses of $10^3$, $10^4$, and $10^5 M_{\odot}$.
Small dots indicate the Log(age)=6.7, 7, 8, 9 and 10 models.
The red arrow represents an $A_{v}=1$ extinction vector (see text). 
}
\label{clusters}
\end{figure}

With an intrinsic distance modulus of $(m-M)_0$=31.4, 
the objects of zone 5 have an absolute bolometric magnitude brighter than -8.4, certainly 
consistent with typical cluster luminosities. Actually, the brightest objects, 
with $I\simeq22$, hence $M_{bol} \leq -9.4$, would be ``populous" clusters, 
according to the definitions by \cite{larsen00} and \cite{billett02}. These 
bigger, more massive clusters are fairly common in
starburst dwarfs, such as NGC~1705 \citep{billett02,annibali09},
NGC~1569 \citep{hunter00,origlia01} and NGC~4449 \citep{annibali11}, and tend to be centrally concentrated, just like these
brightest objects of IZw18. Hence, also if taken as star clusters, the objects 
of zone 5 would confirm a higher  star formation activity in the MB than
in the SB.

To infer the evolutionary properties of unresolved clusters, two photometric
bands are insufficient. Fortunately IZw18 was observed also in the B band with
the WFPC2 by \cite{dufour96b}. We cross-correlated the 31 + 2 objects 
in Zone 5 with the WFPC2 (F435W, F555W) 
photometry presented by \cite{aloisi99}. The cross-correlation 
provides 23 + 2 objects photometrized in F439W, F555W and F814W. 
A visual inspection in our images of the objects missed by the 
cross-correlation 
shows that they are all located in the most crowded MB region,
and thus their position measurement is highly uncertain, also because of 
blending effects.
The WFPC2 magnitudes were transformed into 
Johnson-Cousins $B, V$ and $I$ magnitudes using the transformations by 
\cite{holtzman95}. 

In Fig.~\ref{clusters} we show the 
$B-V$ versus $V-I$ color-color diagram and the $V$ versus $V-I$ color-magnitude 
diagram for zone 5 objects. 
The data are compared with the 
Padova simple stellar populations (SSPs) with metallicity Z$=$0.0004, 
ages from $\sim$1 Myr to $\sim$13 Gyr, and masses of $10^3$, $10^4$, 
and $10^5 M_{\odot}$ \citep{girardi10}.
Zone 5 objects have $V-I$ colors in the range $-0.2$--$1.8$, and $B-V$ colors 
in the range $-0.2$--$1.4$, and are compatible with SSP ages from a few Myr 
to a Hubble time. 
The two objects with $V-I\sim0.4$, $B-V\sim$0.8, 
and $V-I\sim-0.2$, $B-V\sim$0.4, are located in the most crowded region of 
the main body, and therefore their colors suffer larger uncertanties.
The objects concentrated around $V-I\sim0.2$ can be either $\sim100$
Myr old clusters or clusters in the red loop of the models at ages around
$\sim4$ Myr.

We notice that there are some objects with $V-I\ga1$ which are redder than the oldest 
SSPs. We can exclude that they are background galaxies because they are all 
concentrated within IZw18's MB, except for two of them, which are more ``peripheral''.
Another possibility is that these objects are highly reddened young embedded 
clusters; however, in this case they should have $A_V\ga3$ (see reddening 
vector in Figure~\ref{clusters}), 
and thus absolute magnitudes brighter than $M_V\sim-10.75$.
Such high luminosities would imply a super star cluster (SSC) classification
according to the definition by \cite{billett02}. However, it is unlikely 
that IZw18 host so many SSCs, considering that NGC~1569, the dwarf with 
the highest known current SFR, presents only 4 of them \citep{hunter00,origlia01}.
Several studies have discussed stochastic effects on the integrated colors of clusters 
\citep[e.g.][]{silva11,maiz09}. Stochasticity in the IMF sampling of a low-mass cluster can heavily change the 
observed colors with respect to predictions, and in particular, red supergiants make appear the cluster colors 
redder. \cite{cervino04} investigated this issue and derived a luminosity limit under which stochastic effects are 
important. Comparing the luminosities of our clusters redder than $V-I\sim1$ with their predictions for ages older than 100 Myr,
we found that our cluster luminosities are well above that limit.
Therefore, the most likely explanation 
is that these objects are supergiants with ages younger 
than $\sim$20 Myr, implying that the
MB has been forming stars at very recent epochs.

\begin{figure}
\includegraphics[width=8cm]{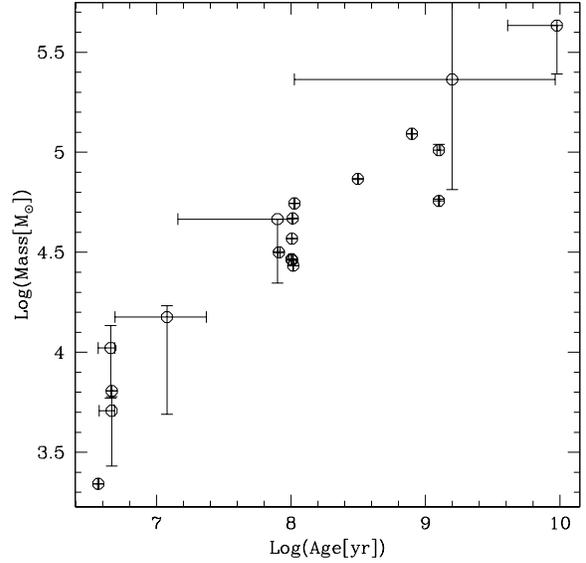}
\caption[]{Masses versus ages derived for the candidate star clusters with the 
Padova SSPs. The objects with $V-I>1$ in Fig.~\ref{clusters} have not been included.
}
\label{clusters_agemass}
\end{figure}

For the zone 5 objects with $V-I<1$ we have inferred ages and masses using the
Padova models. This was done minimizing the difference between model and data colors 
(B$-$V, V$-$I, and  B$-$I) according to a $\chi^2$ criterion, varying the age between $\sim$1 
Myr and $\sim$13 Gyr. We accounted for a Galactic reddening of E(B-V)$=$0.032,  
but did not solve for possible internal extinction, because of the small number of bands (3) and the 
relatively short wavelength range covered by our data. 
Following the prescriptions of \cite{avni}, 
the 1 $\sigma$ errors were derived as the limits of the 0.68 confidence level 
region defined by $\chi^2 < \chi^2_{min} + 1$. 
Cluster masses were obtained from the mass-to-light ratios of the best-fitting models in the I band. 
The results are presented in a mass versus age diagram in Fig.~\ref{clusters_agemass}. 
Cluster ages span the whole Hubble time, even if we notice that the majority of clusters have ages 
younger than a few hundred Myr. Cluster masses range from $2\times10^3 M_{\odot}$ to $4\times10^5 M_{\odot}$. 
The observed correlation between cluster ages and masses is an obvious effect of the 
magnitude limit adopted for zone 5. 

We notice that our analysis provides 5 ``old'' (age$\ga$1 Gyr) clusters, with masses between $6\times10^4$ and 
 $4\times10^5 M_{\odot}$, and  totaling $\sim10^6  M_{\odot}$. This implies that $\sim$3\% of the stellar mass in IZw~18 is in the form 
of old clusters, which sounds unlikely given the fact that this ratio has been found to be considerably lower in other dwarf irregulars with 
evidence for significant SF at ages $>$1 Gyr (e.g., $<$0.3\% in NGC~4449, Annibali et al.~2011).  
This suggests that the majority of the candidate old clusters in IZw18 are instead resolved supergiants. 

Concluding, our analysis seems to suggest that the majority of the objects in zone 5 are indeed individual 
stars, even if we can not exclude that some of them are unresolved stars clusters. Nevertheless, even in this scenario, our 
analysis provides evidence for a particularly active SF over the past hundred Myr. 
 
\section{Discussion of the Results}

The evolutionary properties derived above for the stars measured in IZw18 are 
similar to those of other metal-poor starbursting dwarfs. For instance, once
the difference in distance modulus $\Delta (m-M)_0 \simeq 0.7$ is taken into 
account, magnitudes and colours of the various types of
stars on the red plume of IZw18 are in perfect agreement with those of the
corresponding stars in SBS1415+437, the other very metal-poor BCD whose stellar
populations have been resolved by HST/ACS \citep{aloisi05}. Indeed, the CMD
of SBS1415+437, whose intrinsic distance modulus is 30.66,  shows an
unmistakable TRGB at $I=26.67$ and the horizontal finger typical of carbon
stars at $I=25.78$ \citep{aloisi05}. SBS1415+437 has a red plume more
populated than IZw18, as if it had been more active in previous epochs. 
In turn, the blue plume of IZw18's MB is more prominent than that
of SBS1415+437, suggesting that its current and recent star formation activity
is proportionally higher. 

The results described in the previous sections allow us to 
draw a qualitative picture of the evolution of the MB and SB. 
Both components contain old and young stars in all evolutionary phases,
but it is apparent that the two stellar systems are distinct from each 
other and have had different histories.  
The quantitative star formation history (SFH) of the two bodies of IZw18 
will be derived from the CMDs using the synthetic CMD method 
\citep[see e.g.][and references therein]{cignoni10} and will be presented in a 
forthcoming paper \citep{annibali11}.

\subsection{The Main Body}

The spatial distribution of stars in different evolutionary phases
displayed in Figures~\ref{image814} and \ref{map07}  suggests that the 
MB is indeed one single system and not ``two galaxies separated and 
interconnected by a narrow luminous bridge" as originally thought  
by \cite{zwicky66}. 
We can however understand how he was misled by the 8-shaped distribution 
of the young blue stars.

Young massive stars (zones 1 and 3 of Figure~\ref{cmdmap07}) 
cover the entire MB. Zone 3 stars are concentrated on 
the NW and SE star forming regions, following the traditional scheme of 
starburst dwarfs, where massive young stars usually group together and are
concentrated in the inner regions, although not necessarily at the very 
center of the galaxy. The NW region contains a larger number of bright 
massive zone 3 objects than the SE region. This suggests a more 
recent/higher star formation activity in the NW clump. 

Zone 1 stars seem (see Figure~\ref{map07}) intriguingly 
to avoid the areas where zone 3 stars are located. Is this because they are
different objects? Since they are redder, zone 3 stars might be
more evolved or more metal rich than zone 1 stars. We rather believe that 
either zone 1 stars lie on gas filaments and are slightly blue-shifted by 
residual gas contamination increasing the F555W flux, or that differential 
reddening slightly affects zone 3 objects. Throughout this paper we have
only considered the foreground Galactic reddening, $E(B-V)=0.032$, but IZw18
does contain a lot of gas and presumably dust and molecules, although in
quantities limited by its low metallicity \citep{cannon02}. It is thus
plausible that absorbing interlopers are more abundant  in the innermost, star
forming regions and that they make the stars of those regions appear redder. In
this framework, zone 3 stars are the same kind of objects as zone 1 stars,
only affected by internal extinction of the order of $\Delta E(V-I)\simeq$0.1.

Lower mass young stars (zones 2 and 4) could help in discriminating 
between the above interpretations, but they are faint and are much more 
severely affected by crowding than by any other effect. Indeed, although 
likely to exist also where massive young stars are found, they are only 
detectable where the latter don't outshine them. As discussed in the 
previous sections, in F10's catalogue many of these
objects lie on the gas filaments protruding from the NW spur. While the vast 
majority of these faint F10 detections on the filaments are likely to be gas 
peaks misinterpreted as stars, we do believe that some of the brighter ones, 
which appear as actual stellar sources when individually and visually 
inspected in the images, are indeed very young stars or unresolved star 
clusters \citep[see also][]{chandar05}. This leads to speculate that star 
formation is occurring at least in the portions of the filaments closest 
to the MB. Unfortunately, we 
don't know whether these filaments are only made of ionized gas or also of 
neutral one, since the current resolution of HI data (F. Fraternali private 
communication, Lelli et al. in preparation) doesn't allow one to resolve them. 
Neither do we know whether  the filaments are infalling, outflowing  or 
comoving with respect to the MB. From van Zee's et al. (1998, their figure 6) 
comparison of the $H_{\alpha}$ and HI  distributions, the filaments appear to 
be a distinct feature of the ionized gas, but the HI resolution is definitely 
insufficient.  It is thus difficult to assess if star formation is physically 
plausible there; certainly it is extremely unlikely that it occurs in hot, 
possibly outflowing bubbles, such as many of IZw18 ionized filaments 
\citep{martin96}.

Also the massive objects of zone 5 are centrally concentrated, as usually found
in starburst dwarfs. Whether they are resolved supergiants or unresolved star 
clusters, they are the signature of strong star formation activity. The
evidence that they are more abundant in the NW than in the SE clump suggests 
that the activity in the former is higher than in the latter.
The NW clump is also known to contain Wolf-Rayet stars 
\citep[see e.g.][]{hunter95,izotov98}. Yet, from emission line spectroscopy, 
\cite{izotov98} suggested it to be not as young as the SE clump: 10 Myr 
the former and 5 Myr the latter (and 20 Myr the SB). The CMDs of 
Figure~\ref{mb_separate} do not seem to confirm their suggestion. Rather, our 
CMDs suggest that the star formation activity in the NW is more recent than in 
the SE. The circumstance that, of the variable stars reliably characterized by 
F10, the two massive ULP  Cepheids V1 and V15 are located in the NW region, 
while the lower mass Classical Cepheid V6 and the fainter  LPV V7 are located 
in the SE region, favours the view that the NW on average contains younger 
stars than the SE.

All the lower mass evolved stars are likely to be evenly distributed in the MB.
RGB stars seem to be located more externally than AGBs, but this is 
most probably a selection effect. 
We ascribe their apparent confinement in the outer regions 
to the severe central crowding which prevents their detections. Similarly, we 
interpret their asymmetrical horseshoe distribution as due to the extinction 
from the thick gas filaments of the northern areas 
\citep[see][for the spatial distribution of dust]{cannon02}. 
There is no doubt that faint objects are more easily detected in the 
SE external portions of the MB.

The resulting overall scenario of the MB consists of a system with two active,
distinct regions of current star formation, one (the NW one) stronger 
and younger than the other. These two active clumps are embedded in a fairly 
homogeneous body of stars of all ages, from a few
Myr to several (possibly 13) Gyr. 

\subsection{The Secondary Body}

The SB is in a rather different situation. As mentioned above, the 
weaker emission lines of its spectrum already led 
\cite{izotov98} to suggest that it is 
older than the MB emission regions.  WFPC2 photometry allowed \cite{aloisi99} 
to quantitatively assess that its recent SFR is much lower than in 
the MB. The ACS photometry confirms these results. 

Our CMDs show that the SB does contain young blue plume stars, but they 
are more likely stars at least 10 Myr old on the blue loop phase. Stars more 
massive than $\sim15 M_{\odot}$ do not seem to be still alive. The SB also 
appears to contain much less ionized gas than the MB: a further 
confirmation of its low recent activity. 

On the other hand, low and intermediate mass stars are at least as 
numerous in the SB as in the MB, with the objects of zone 4 (low mass MS 
stars) even outnumbering those in the MB (see Table~1). This is a strong 
indication that at epochs earlier than $\sim$100 Myr ago, the SB must 
have been actively forming stars as, or more than, the 
MB\footnote{A deeper and more detailed analysis of the relative star formation 
activity in both bodies, taking into acount completeness will be present in 
\citet{annibali11}.}.
If we compare the CMD of the MB SE region (Figure~\ref{mb_separate}) with 
that of the SB (Figure~\ref{cmdisotracks}), we see that the two red plumes are 
similar, both in the RGB and the AGB sequences. The blue plume, although  
containing similar numbers of stars, clearly shows that those in the SB are 
more evolved than those in the SE, that reach brighter magnitudes and bluer 
colours. The AGB and the RGB of the SB are well defined. The colour 
extension of the latter, combined with the small metallicity range of 
the system, indicates that stars with the oldest possible ages (13 Gyr) 
are present. 

We conclude that the SB has started its star formation activity up to $\sim13$ 
Gyr ago, and has remained active since then, at rates comparable with 
those in the MB until relatively recently. However, in the most recent 
epochs (i.e. later than 10 Myr ago) its SFR has dropped to values 
significantly lower than in the MB.

\section{Summary and conclusions}

We have studied the evolutionary properties and spatial distribution of the
stars resolved in IZw18 by HST/ACS proprietary and archival photometry.  
To this aim we have analysed both the Al07 and the F10 catalogues and 
found consistent results, independently of the different selection 
criteria adopted in the two approaches. Nonetheless, to be 
conservative, we have based our conclusions on the Al07 most 
restrictively selected catalogue.

The comparison of the CMDs resulting from the ACS photometry with stellar 
evolution models (tracks and isochrones) indicates that stars of all 
ages are present in the two bodies. In spite of the impossibility of 
reaching the oldest MS turn-off in a galaxy 19 Mpc away, the extremely 
low metallicity of IZw18 relaxes the age-metallicity degeneracy and makes 
the red edge of the RGB a signature of the presence of stars up to possibly
13 Gyr old.  The oldest detectable stars are best visible in the SB and the 
SE portion of the MB, where crowding is less severe, but are present also 
in the rest of the MB, although measured with larger uncertainties.

We have found that the stars are homogeneously distributed over the two  
bodies, with the younger ones more centrally concentrated, as always 
found in starburst dwarfs, and the older/fainter ones spread out to the 
system periphery, also because of selection effects in their detectability.
From the maps of the stars spatial distribution (Figure~\ref{map07}) we 
conclude that old and intermediate age stars are actually distributed 
homogeneously over the two bodies.

Summarizing the results presented by Al07 and F10 and those obtained here, 
the main properties of IZw18 stellar populations are:
\begin{itemize}
\item The distance independently estimated from the TRGB (Al07) and the 
Cepheids (F10) is $\sim19$ Mpc.
\item As far as stars are concerned, the SB is completely separated from 
the MB: no star is detected between the two bodies. If tidal interactions 
are at work, they don't appreciably affect the 
distribution of the
stellar populations of the two components.
\item IZw18  contains stars in all the 
evolutionary phases visible at its distance: massive and intermediate 
mass stars on the MS, blue and red supergiants, AGB and TP-AGB stars, 
carbon stars, RGB stars. Hence, it hosts stars of all ages, from a few Myr 
up to possibly 13 Gyr. 
\item The MB hosts two bright star forming regions (NW and SE), embedded 
in a common and homogeneous environment of older stars. Some stars may 
have been recently formed on the gas filaments protruding from the NW 
portion of the MB.
\item While old and intermediate age stars are present in similar amounts 
in the MB and in the SB, the youngest stars are only found in the MB and 
mostly concentrated in its NW region. The SE region has also been
active recently, but at a lower current rate; the SB has had definitely 
less star formation than the MB in the last tens of Myr. 
\item Of the 34 candidate variable stars studied by F10, 19 and 15 are in, 
or around,  the MB and the SB respectively. All the confirmed ones are 
however in the MB: the two ULP Cepheids in the NW region;  the Classical 
Cepheid and the two LPVs in the SE region.
\item At IZw18's distance, star clusters are not distinguishable from 
point-like sources. It is thus possible that some of the brightest 
objects in the main body are actually clusters and not individual stars. 
Comparing their colors with simple stellar population models, we find 
that the candidate clusters would span ages from a few Myr to a Hubble 
time, with the majority of them being younger than $\sim$200 Myr. 
However, the fraction of stellar mass in old (age$\ga$1 Gyr) clusters would be 
significantly higher than what found in other dwarfs, suggesting that 
the majority of the candidate old clusters are instead resolved stars. 
The reddest ($V-I>1$)  brightest objects cannot be unresolved clusters and must be 
supergiants with ages younger than $\sim$20 Myr.

\end{itemize}

These results show that IZw18 is not unique in the sky, but shares 
the evolutionary properties typical of BCDs and, more in general, of 
late-type dwarfs \citep[see][for a review]{tolstoy09}. It is not 
experiencing now its first burst of star formation, but has been forming 
stars over many Gyrs, most likely over the whole Hubble time. 
This implies that its extremely low metallicity cannot be  explained by a 
low chemical enrichment due to extremely recent star formation. At best, 
one may hope that the SFR has been low until relatively recently 
\citep[see e.g.][]{legrand00}, although the conspicuous presence of stars 
of all ages does not make this likely.  

The same problem obviously concerns all the other metal-poor BCDs where 
stars many Gyr old have been found, like SBS1415+437 \citep{aloisi05}. We 
recall that no genuinely young galaxy has been found yet by anybody.  All 
the galaxies whose stellar content has been resolved so far contain  stars 
as old as the lookback time allowed by the available instrumentation, 
whatever their morphological type and metallicity 
\citep[see e.g.][and references therein]{tosi09,cignoni10}.  
Hence, since stellar nucleosynthesis has had plenty of time to pollute 
them, the only ways to explain the low observed metallicity of these 
galaxies are either to dilute sufficiently their insterstellar medium 
by accretion of primordial or very metal poor gas, or to remove the 
stellar nucleosynthesis products through galactic winds triggered by 
supernova explosions, or both. Chemical evolution models of late-type 
dwarfs and BCDs have shown long ago 
\citep{matteucci85,pilyugin93,marconi94} that infall of metal poor gas 
alone is not sufficient to reproduce the low metallicity of starburst 
dwarfs, and that galactic winds must be invoked too.

Observational evidence for galactic winds has been found for a handful 
of starburst dwarfs \citep{bomans07,grimes09}, and IZw18 is indeed one of 
those where the outflowing (Martin 1996) ionized gas associated with the 
most recent SF activity may indeed have a speed higher than the escape 
velocity and be able to remove metals from the system. 
It would be crucial to get detailed information also on the cold gas,  
see how it is spatially distributed,  and what its kinematics is.

Understanding the chemical and dynamical evolution of IZw18 represents a 
key step to approach the evolution of all dwarf galaxies as well as that 
of more massive systems which may have formed from the assembly of small 
objects like it.  

\acknowledgments
We thank Filippo Fraternali and Renzo Sancisi for interesting conversations on 
the HI gas properties. Financial support for this study was provided by ASI
through contracts COFIS ASI-INAF I/016/07/0 and ASI-INAF I/009/10/0. Support 
to the US co-authors was provided by NASA through 
grants associated with program GO-10586 from the Space Telescope Science
Institute (STScI), which is operated by the Association of
Universities for Research in Astronomy, Inc., under NASA contract
NAS5-26555.

\clearpage

\end{document}